\documentclass[12pt,a4paper]{article}
\usepackage[utf8]{inputenc} 
\usepackage[a4paper,top=2.5cm,bottom=2.5cm,left=2cm,right=2cm]{geometry}
\usepackage[left]{lineno}

\usepackage{authblk}

\usepackage{footmisc}
\usepackage{amsfonts}

\usepackage{lineno}

\usepackage{adjustbox}
\usepackage{graphicx}
\usepackage{ulem}
\usepackage{booktabs}
\usepackage{hyperref}       
\hypersetup{
    colorlinks,
    citecolor=black,
    filecolor=black,
    linkcolor=black,
    urlcolor=black
}
\usepackage{amsmath}
\usepackage[dvipsnames]{xcolor}
\usepackage{xcolor}
\definecolor{lightblue}{RGB}{40, 100, 200}

\usepackage{titlesec}
\titlespacing*{\paragraph}{0pt}{*0}{1em} 
\titleformat{\paragraph}[runin]        
  {\bfseries\itshape}                
  {}                                    
  {0pt}                                 
  {}    
\usepackage{titlesec}
\usepackage{mathtools}
\DeclarePairedDelimiter\ket{\lvert}{\rangle}

\newcommand{\hsb}{\Ham_{SB}}

\newcommand{\ada}{\hat{a}^{\dagger}\hat{a}}

\newcommand{\pz}{\hat{\sigma}^{z}}
\newcommand{\px}{\hat{\sigma}^{x}}

\newcommand{\pmi}{\hat{\sigma}^{-}}
\newcommand{\ppl}{\hat{\sigma}^{+}}
\newcommand{\opa}{\hat{a}}
\newcommand{\opad}{\hat{a}^{\dagger}}

\newcommand{\Comment}[1]{}
 
\newcommand{\RS}{\hat{\mathcal{S}}}
\newcommand{\RSbar}{\tilde{\mathcal{S}}} 
\newcommand{\Loop}{\mathcal{L}} 
\newcommand{\Uop}{\hat{\mathcal{U}}} 
\newcommand{\Ham}{\hat{\mathcal{H}}}

\title{\huge Loop Quantum Photonic Chip for Coherent Multi-Time-Step Evolution}

\begin{document}

 \setlength{\parindent}{0pt}
 \setlength{\parskip}{6pt}
 \setlength{\columnsep}{15pt}
 \bibliographystyle{unsrt}

\author[1,2]{Yuancheng Zhan}
\author[3,4]{Hui Zhang\protect\footnotemark[1]}
\author[5]{Rebecca Erbanni}
\author[6]{Andreas Burger}
\author[1]{Lingxiao Wan}
\author[1]{Xudong Jiang}
\author[1]{Sanghoon Chae}
\author[7]{Ai-Qun Liu\protect\footnotemark[2]}
\author[5,2,8,9]{Dario Poletti\protect\footnotemark[3]}
\author[1,2,9,10]{Leong Chuan Kwek\protect\footnotemark[4]}

\footnotetext[1]{Corresponding author: jovie\_huizhang@tongji.edu.cn}
\footnotetext[2]{Corresponding author: aiqun.liu@polyu.edu.hk}
\footnotetext[3]{Corresponding author: dario\_poletti@sutd.edu.sg}
\footnotetext[4]{Corresponding author: cqtklc@nus.edu.sg}
\affil[1]{School of Electrical and Electronic Engineering (EEE), Nanyang Technological University, Singapore 639798, Singapore}
\affil[2]{Centre for Quantum Technologies (CQT), National University of Singapore, Singapore 117543, Singapore}
\affil[3]{Institute of Precision Optical Engineering, School of Physics Science and Engineering, Tongji University, Shanghai 200092, China}
\affil[4]{MOE Key Laboratory of Advanced Micro-Structured Materials, Shanghai Institute of Intelligent Science and Technology, and Shanghai Frontiers Science Center of Digital Optics, Tongji University, Shanghai 200092, China}
\affil[5]{Science, Mathematics and Technology Cluster, Singapore University of Technology and Design, 487372, Singapore}
\affil[6]{Department of Computer Science, University of Toronto, Toronto, Canada }
\affil[7]{Institute of Quantum Technologies (IQT), The Hong Kong Polytechnic University, Hong Kong, China}
\affil[8]{Engineering Product Development Pillar, Singapore University of Technology and Design, 487372 Singapore}
\affil[9]{MajuLab, CNRS-UNS-NUS-NTU International Joint Research Unit, Singapore 117543, Singapore}
\affil[10]{National Institute of Education, Nanyang Technological University, Singapore 637616, Singapore}

\date{}
\vspace{-3em}
\maketitle
\vspace{-3em}
\begin{abstract}
Quantum evolution is crucial for the understanding of complex quantum systems. However, current implementations of time evolution on quantum photonic platforms are limited by both low photon generation efficiency and high propagation loss, making photon detection difficult. Furthermore, the single-layer complexity of most implementations cannot support multi-step quantum simulations. In this work, we present a {\it loop quantum photonic chip} (Loop-QPC) designed to efficiently simulate quantum dynamics over multiple time steps in a single chip. Our approach employs a recirculating loop structure to reuse computational resources and eliminate the need for multiple quantum tomography steps or chip reconfigurations. A cycle-or-measure circuit at the input and output stages allows dynamic routing of photons between evolution and measurement, enabling efficient control over the simulation loop and minimizing loss. We experimentally demonstrate the dynamics of the spin-boson model on a low-loss Silicon Nitride integrated photonic chip. The Loop-QPC achieves a three-step unitary evolution closely matching the theoretical predictions. These results establish the Loop-QPC as a promising method for efficient and scalable quantum simulation, advancing the development of quantum simulation on programmable photonic circuits. 
\end{abstract}

\section{Introduction}
Simulating the evolution of quantum systems~\cite{mcfadden2002quantum,motta2020determining} represents one of the most promising applications of programmable quantum processors~\cite{daley2022practical,endo2020variational}, providing practical means to explore and understand complex quantum systems that remain intractable for classical computers. A notable testbed for studying complex quantum evolution is the spin-boson model~\cite{leggett1987dynamics,dolgitzer2021dynamical} that describes the dynamics of a spin coupled to a bath. This model plays a crucial role in understanding light-matter interactions~\cite{frisk2019ultrastrong}, quantum chemistry~\cite{mcardle2020quantum}, charge transfer \cite{CukierMorillo1989}, exciton transport \cite{HuelgaPlenio2013}, macroscopic quantum tunneling in superconducting systems \cite{Han1991}, polarons \cite{weiss2012}, and the nonadiabatic dynamics of molecules~\cite{ollitrault2020nonadiabatic}. 

Recently, there has been significant progress in the development of different quantum processor platforms, including ultracold gases \cite{langen2024quantum}, trapped ions \cite{moses2023race}, superconducting qubits \cite{cao2023generation}, and photonics \cite{o2009photonic,lib2024resource}, even for the simulation of spin-boson-like systems \cite{sun2025quantum,so2024trapped,tang2024simulating}. 
The development of photonic systems has also led to on-chip quantum processors ~\cite{yin2025experimental,wang2020integrated,zhang2021optical,zhang2023encoding}, solid-state solutions which work at room temperature, are reprogrammable, and whose realization is compatible with existing semiconductor fabrication technology. Importantly, they can be used for fault-tolerant computing, variational approaches, and analogue quantum simulation~\cite{SlussarenkoPryde2018}. Integrated photonic chips naturally implement unitary evolution through spatial and temporal interference of optical modes~\cite{crespi2013anderson,harris2017quantum,sparrow2018simulating}, which allows scalable and parallel simulation of quantum Hamiltonians. They further offer low loss, high bandwidth, and stable control of large-scale interferometers, making them well-suited for programmable linear-optical quantum processors~\cite{harris2018linear,pelucchi2022potential,chi2023high}. Recent works have demonstrated programmable photonic processors with 15 to 128 modes~\cite{wang2018multidimensional,taballione202320,bao2023very,hua2025integrated,ahmed2025universal}, highlighting steady progress toward scalable linear optical platforms. These advances suggest that large-scale simulations on photonic chips may be viable in the near term.   

However, implementing quantum evolution on photonic platforms still presents significant challenges. Current approaches primarily rely on sequentially modifying chip parameters to perform time evolution for different time durations ~\cite{sparrow2018simulating}. This requires recalculating the unitary transformations~\cite{clements2016optimal,bogaerts2020programmable} at each step and repeatedly loading these parameters onto the chip. These introduce computational overhead from frequent recalculation of unitary parameters, and necessitate repeated digital-to-analog and analog-to-digital conversions between the photonic hardware and digital controllers. 
Alternatively, quantum tomography-based methods~\cite{wang2020integrated} measure the output state at each time step and reload this state as the subsequent input, which is not a scalable solution, and it is also not solved by multiplexing techniques such as wavelength-division multiplexing \cite{elshaari2017chip} and time-division multiplexing \cite{bacco2017space,harris2014integrated}.

\begin{figure*}[ht!]
\centering
\includegraphics[width=1\columnwidth]{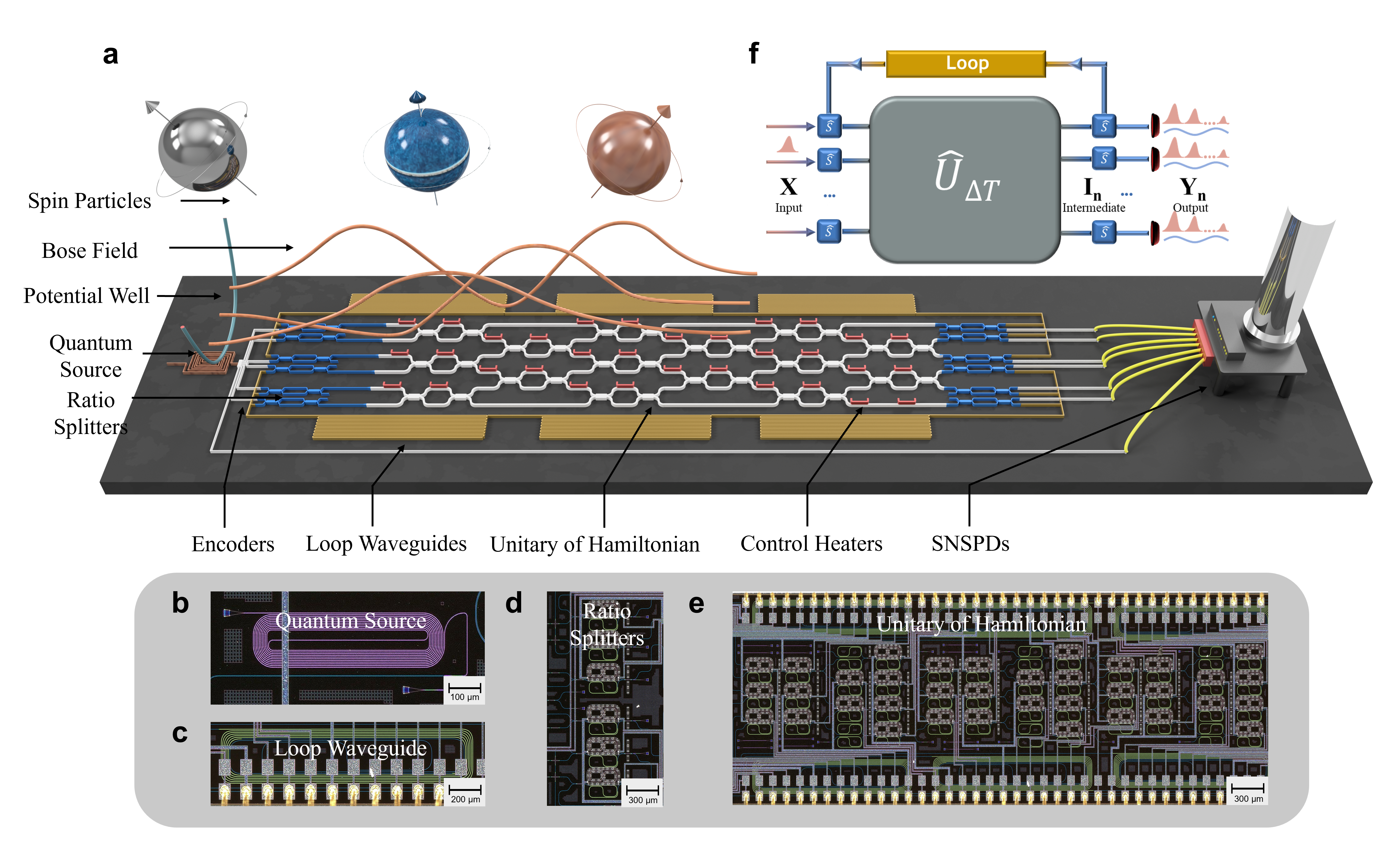}
\caption{Overview of the Loop-QPC architecture. (a) Schematic illustration of the spin-boson model (top), implemented using our loop quantum photonic chip (bottom). (b)-(e) A more detailed representation of the Loop-QPC architecture with actual microscope images of key components: (b) the quantum photon-pair source, (c) the integrated loop waveguide, (d) the ratio splitters for conditional routing, and (e) the unitary circuit implementing Hamiltonian dynamics. (f) The flow chart of the  Loop-QPC. The input state $\boldsymbol{X}$ goes through ratio splitters and unitary to become the intermediate stage, $\boldsymbol{I}_{n}$, which can be looped back towards the unitary, or exit as $\boldsymbol{Y}_{n}$ towards the measurement module. 
}%
\label{fig:1}
\end{figure*}

To overcome these limitations, we propose a loop quantum photonic chip (Loop-QPC) that enables efficient quantum evolution by reusing a single chip for multiple time steps. The core of our design lies in a recirculating loop structure with the cycle-or-measure circuit. This circuit determines whether a photon continues through additional evolution cycles or is directed to the measurement stage. It enables controlled simulation of multiple time steps within a single experimental run, thereby eliminating the need for both quantum state tomography and repeated parameter reconfiguration.
We thus program our Loop-QPC to execute the quantum evolution of the spin-boson model for time step $\Delta T$. By leveraging the low-loss properties of Silicon Nitride (SiN) and the high detection efficiency of Superconducting Nanowire Single Photon Detectors (SNSPDs), we are able to use our photonic chip to simulate the quantum evolution for three distinct time steps with high fidelity, while being able to take measurements at each of the time steps.

\section{Model}
We consider the spin-boson model with a single spin coupled to a harmonic oscillator as depicted at the top of Fig.~\ref{fig:1}(a). In the strong coupling regime, its evolution in time is given by the Hamiltonian \cite{thorwart2004dynamics},
\begin{equation} \label{eq:spinboson}
\hsb = \hbar\omega \ada + \frac{1}{2} (h\pz + \epsilon \px ) + \lambda \px \ (\opad + \opa),  
\end{equation}  
where $\hbar$ is Planck's constant, $\opad$ and $\opa$ are the creation and annihilation operators for the harmonic oscillator, respectively, while the operators $\px=\ppl+\pmi$ and $\pz$ are Pauli operators acting on the spin. The parameter $h$ represents the local magnetic field in the $z-$direction, while $\epsilon$ is the field in the $x-$direction. The coupling strength between the spins and the harmonic oscillator is denoted by $\lambda$, and $\omega$ is the frequency of the oscillator.  

To experimentally simulate the unitary time evolution generated by $\Ham_{SB}$ over a time $T$, we decompose the evolution over $N_T$ time  steps of duration $\Delta T = T/N_T$ as 
\begin{align}   
\Uop_T = \left[\Uop_{\Delta T}\right]^{N_T} = \left[e^{-i \hsb \Delta T/\hbar}\right]^{N_T},       
\label{eq:unitary}
\end{align} 
where $\Uop_{\Delta T}$ is the evolution operator for an arbitrary $\Delta T$. Here, since $\Ham_{SB}$ is static, the same unitary is applied repeatedly, which allows implementation on photonic hardware without requiring time-dependent control.

To implement the spin and bosonic operators in $\hsb$, we map them to Pauli operators using unary encoding. Unary encoding \cite{ramos2021quantum,zhang2021efficient}, also known as `one-hot' encoding, assigns each state to a distinct channel, i.e., for a system of dimension $N$, the $i-$th state is given by $(0_1 ... 1_i ... 0_N)$. This scheme produces deterministic outcomes and allows for the realization of entangling operations using only linear optics~\cite{stanisic2017generating,zhang2023efficient}. To encode our model on a photonic chip, each basis state is mapped to a specific channel, whereby a single photon in a specific channel with no photons in other channels represents a particular basis state. 
To represent the spin-boson model state and its evolution, we consider both levels for the spin$-1/2$ ($|\downarrow\rangle_s$ and $|\uparrow\rangle_s$) and the first three levels for the (weakly occupied) harmonic oscillator ($|0\rangle_b$, $|1\rangle_b$, and $|2\rangle_b$). This yields a total Hilbert space dimension of 6, requiring 6 optical channels for unary encoding, consistent with the approach used in Ref.~\cite{burger2022digital}, with each basis state encoded in a distinct channel on our photonic chip. For example, for the initial state, we consider the spin in the excited state $|\uparrow\rangle_s$ and the bosonic mode initialized in a single-photon Fock state $|1\rangle_b$, resulting in the composite state $|\uparrow\rangle_s \otimes |1\rangle_b$. Under unary encoding, this state is mapped to a 6-dimensional vector as $|0_1, 1_2, 0_3, ..., 0_6\rangle$, where each subscript corresponds to an optical channel index.

To connect our unary-encoded demonstration to a multi-photon interference regime, we consider an 8-dimensional spin–boson model, which can be encoded using three qubits via a dual-rail scheme. Each qubit is represented by a single photon occupying one of a designated mode pair: $(1,2)$, $(3,4)$, and $(5,6)$.
Injecting three indistinguishable photons into the $6 \times 6$ unitary $W$ defines a computational subspace spanned by the eight logical basis states, $s_1=(1,3,5),\;  \ldots,\; s_8=(2,4,6)$, with each $\ket{s_j}$ mapped to a three-photon Fock state. After propagation, we post-select output events with exactly one photon in each mode pair, projecting the dynamics onto an effective $8\times8$ logical unitary $\tilde{U}$ with elements,
\begin{equation}
\tilde U_{ij}
=
\mathrm{perm}\!\big(W[s_i,s_j]\big)
=
\mathrm{perm}\!\left(
\begin{bmatrix}
w_{s_i(1),s_j(1)} & w_{s_i(1),s_j(2)} & w_{s_i(1),s_j(3)}\\
w_{s_i(2),s_j(1)} & w_{s_i(2),s_j(2)} & w_{s_i(2),s_j(3)}\\
w_{s_i(3),s_j(1)} & w_{s_i(3),s_j(2)} & w_{s_i(3),s_j(3)}
\end{bmatrix}
\right),
\end{equation}
where $\mathrm{perm}(\cdot)$ denotes the matrix permanent.
This construction lifts the evolution from single-particle matrix elements to multi-photon permanents, enhancing interference complexity in principle.
Experimental limitations remain, and further details are provided in Supp. Mat. A.
\section{Loop-QPC Design}
Our proposed Loop-QPC architecture, illustrated in Fig.~\ref{fig:1}(a), consists of the following main modules: 
The quantum photon generation module (Fig.~\ref{fig:1}(b)), which produces signal and idler photon pairs for computation and correlation measurement, respectively.
The data encoding module is responsible for initializing the input quantum state.
The loop module $\Loop$ (Fig.~\ref{fig:1}(c)), enabling multiple traversals of photons through the unitary circuit for multi-step evolution.
The cycle-or-measure modules (Fig.~\ref{fig:1}(d)) at both the input and output stages of the chip are each composed of six MZIs, with each MZI containing one phase shifter and one amplitude modulator. These elements enable control over the splitting ratio between looping and measurement paths. This routing mechanism ensures that photons can be directed either into the next evolution cycle or to the output detectors as required.
The unitary quantum computation module $\Uop_{\Delta T}$ (Fig.~\ref{fig:1}(e)), which implements the quantum evolution over each time step.
The control module, which programs unitary parameters via on-chip MZI heaters.
The measurement module, utilizing SNSPDs for photon detection and coincidence counting

Here we describe the evolution of the quantum state in terms of three discrete stages, illustrated in Figs.~\ref{fig:1}(f): the input stage $\boldsymbol{X}$ corresponding to the initial input quantum state in the Loop-QPC, intermediate stages $\boldsymbol{I}_n$ representing the signal after going through the unitary portion $n$ times, and the output stage $\boldsymbol{Y}_n$ for the measurement devices after passing $n$ times through the unitary. We denote the ratio splitters as $\RS_\mu$ and $\RSbar_\mu$, where $\mu=i,o$ indicates whether the ratio splitter is at the input ($i$) or output ($o$) of the unitary module, while the hat or tilde stands for different directions.
We can thus write   
\begin{equation}
\begin{aligned}
&\boldsymbol{I}_{1}  =\Uop_{\Delta T} \cdot \RS_i \cdot \boldsymbol{X} \\
&\boldsymbol{I}_{n}  =\Uop_{\Delta T} \cdot \RSbar_i \cdot \Loop \cdot \RSbar_o \cdot \boldsymbol{I}_{n-1} \\
&\boldsymbol{Y}_{n}  = \RS_o \cdot \boldsymbol{I}_{n}, \quad n=1,2,\cdots,N_T. 
\end{aligned}\label{eq:chipeq}
\end{equation} 

The implementation of an arbitrary 6-mode unitary $\Uop_{\Delta T}$ consists of 15 unit cells, each constructed by a Mach-Zehnder interferometer (MZI) composed of two symmetric (50/50) beam splitters and an internal/external relative phase shift $\theta$ and $\phi$. 
The transformation implemented by each MZI is mathematically described as,
\begin{equation}\label{eq:mzi}
\mathcal{T}=\left[\begin{array}{cc}
e^{i \phi} \cos \theta & -\sin \theta \\
e^{i \phi} \sin \theta & \cos \theta
\end{array}\right].
\end{equation}
The initial values for $\theta$ and $\phi$ are computed first using Clements et al. decomposition \cite{clements2016optimal} (see details in Supp. Mat. B) and further refined through machine learning techniques (see details in Supp. Mat. C) to enhance the fidelity.
Our compact phase shifter design shortens each MZI unit cell shown in Fig.~\ref{fig:1}(e), effectively reducing the total propagation loss. 
Thanks to high-quality SiN fabrication, the waveguides achieve an optical transmission ($\alpha$) around 0.6 dB/cm. The MZI architecture also facilitates the input/output ratio splitting.

For the experimental realization, we employ an integrated quantum photonic architecture implemented on the chip. 
This architecture minimizes the propagation path length and ensures uniform propagation loss across all paths. 
The loop module incorporates a low-loss delay line in a spiral structure,  designed to ensure a fixed temporal separation between photons from adjacent loops, thereby preventing interference between different loop photons. To achieve this, the loop length must exceed a critical threshold of $\approx 4$cm computed from the speed of light in SiN and the temporal resolution of the high-efficiency SNSPD used for detecting photons. Four primary types of losses are considered: (1) Splitting loss caused by light splitting in ratio splitters at both the input and output of the loop; (2) Propagation loss which includes losses during forward propagation through the QPC $l_{chip}$, and (3) during backward propagation through the delay loop loss $l_{loop}$; (4) other losses, such as coupling loss, detection loss, and off-chip propagation loss, are collectively referred to together as $l_{others}$, for simplicity. Therefore, also following \cite{clements2016optimal, siew2021Review}, the total loss $L$ after $n$ time steps of the Loop-QPC is expressed as 
\begin{equation}
\begin{aligned}
    L(n) &= \RSbar_i \cdot [\RS_i \cdot l_{chip} \cdot l_{loop} \cdot \RS_o]^{n-1} \cdot l_{chip} \cdot \RSbar_o \cdot l_{others} \\
    \text{subject to} & \quad \RSbar_i+\RS_i=1, \quad \RSbar_o+\RS_o=1.  
\end{aligned}\label{eq:loss_theory}
\end{equation}

For $n=3$ in our system, the cumulative output photon loss in Eq.~\ref{eq:loss_theory} is minimized by optimizing the cycle-or-measure modules. To determine the optimal configuration of the MZIs, we minimize the total output loss $L(n)$ by calculating the partial derivatives with respect to $\RS_i$ and $\RS_o$. Let $c= [l_{chip} \cdot l_{loop} ]^{n-1} \cdot l_{chip} \cdot l_{others}$, and the partial derivative of the loss function is,
\begin{equation}
\frac{\partial L}{\partial \RS_i}=\frac{\partial}{\partial \RS_i}\left[\left(1-\RS_i\right) \cdot\left(\RS_i \cdot \RS_o\right)^2  \cdot\left(1-\RS_o\right)\cdot c\right]
\end{equation}
Setting $\frac{\partial L}{\partial \RS_i}=0$, the optimal solution is found to be $\RS_i=2/3 \hat{I}$ and $\RSbar_i=1/3 \hat{I}$, where $\hat{I}$ is the identity. Similarly, we obtain $\RS_o=2/3 \hat{I}$ and $\RSbar_o=1/3 \hat{I}$. This setup ensures that approximately two-thirds of the photons are redirected into the loop at each time step, while one-third proceed to the output, achieving a balanced trade-off between loop depth and output photon signal strength.

\begin{figure}[h!]
\centering
\includegraphics[width=0.9\columnwidth]{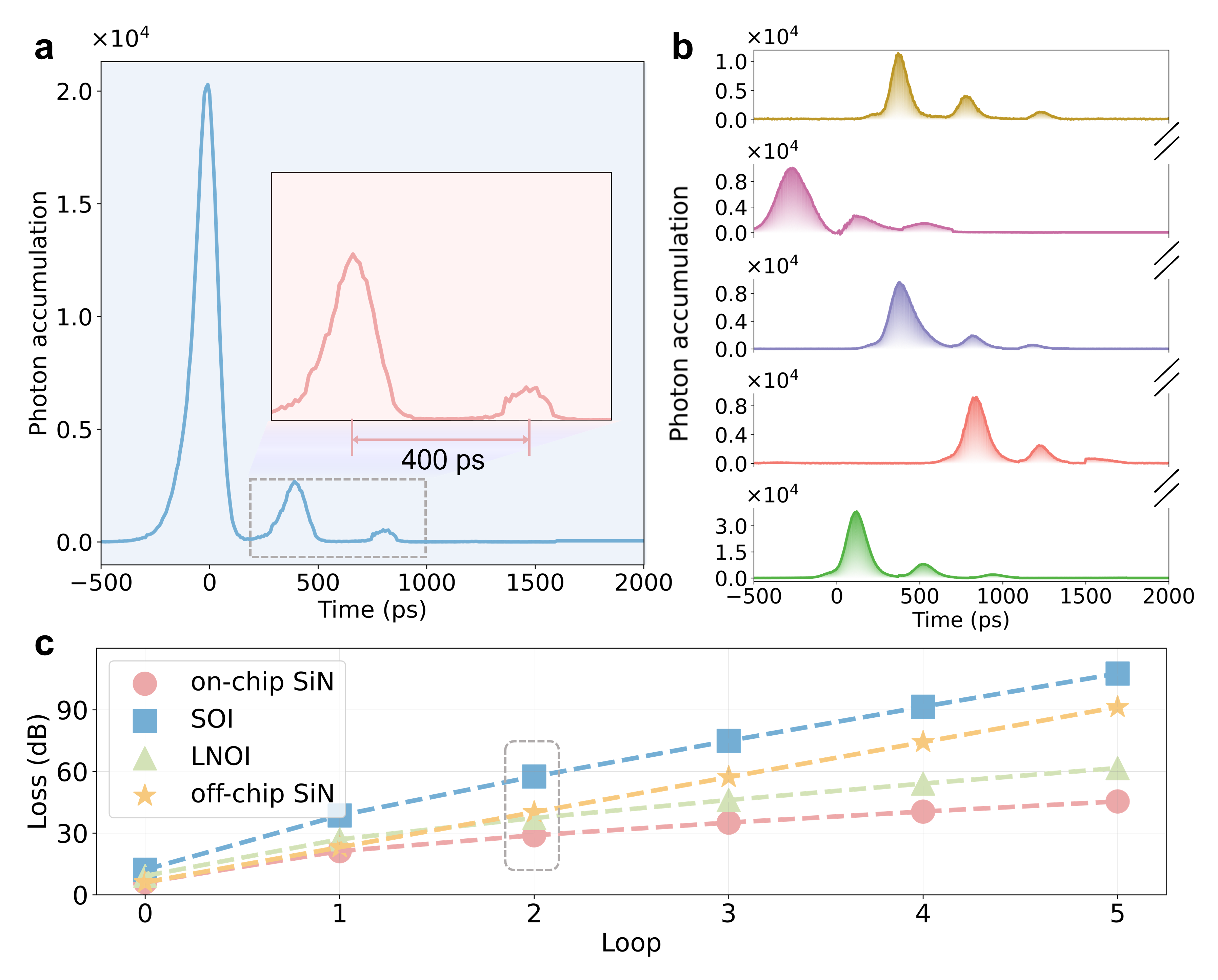}
\caption{Photon accumulation from the first channel (a), and channels 2 to 6 from top to bottom in panel (b). The time zero is set by the coincidence with the idler photon. Inset of panel (a): zoom-in on peak separation of $\approx 400$ ps. (c) Loss in dB of photons versus the number of loops for different circuit designs. In all platforms, the unitary evolution is the identity matrix.}
\label{fig:2}
\end{figure}

The whole Loop-QPC computing process is as follows: a pump laser operating at a 500 MHz repetition rate generates input light, resulting in signal peaks at 2 ns intervals. The pump light with a central wavelength of around 1550 nm initiates a pair of signal ($\approx$ 1545 nm) and idler ($\approx$ 1555 nm) photons in the long waveguide via the spontaneous four-wave mixing process. By regulating the pump power, we balance the photon generation rate and the probability of generating multi-pair terms.  
The signal photons are used for computation, while the idler photons are used as a herald to mark the timing of the signal photon. 
Output photons are coupled to fibers through a grating array, and photon detection is performed using seven ports of SNSPDs, with six dedicated to signal photons and one for idler photons.

\section{Resolution of the Time Steps}
We first resolve three different intensity peaks in our setup for a trivial Hamiltonian described by an identity matrix. Running the experiment for a pair of signal and idler photons produces the measured pair at a rate of $10^4$ photons per second. We record the time of arrival of signal photons in each of the channels, storing these times relative to the arrival time of their corresponding idler photon.    
We depict the count of photons per time and channel. Figure~\ref{fig:2}(a) displays the detailed count histogram for channel 1, while Fig.~\ref{fig:2}(b) shows the histograms for channels 2 through 6 (from top to bottom). These results show a clear separation between three different arrival times, after the chip is passed through one, two, or three times.  
The peaks are well above background noise and do not overlap, confirming the Loop-QPC's ability to maintain clear signal separation. 

To stress the advantages of the Loop-QPC we have designed, we numerically compare the photon losses versus the number of loops for different designs. 
We compare SOI platforms~\cite{zhang2021optical}, LNOI platforms~\cite{sund2023high}, and off-chip loop-based SiN setups~\cite{chen2022iterative}. In our loss calculations, we considered only the difference in transmission losses and functional units loss to each platform, while keeping all other conditions, such as the photon wavelength (centered at 1550 nm for all platforms), consistent with our own experimental setup. For the design with the off-chip loop case, we also account for the photon coupling loss during each loop. As shown in Fig.~\ref{fig:2}(c), our Loop-QPC design has reduced losses because it avoids some of the key issues of the other technologies, i.e., SOI platforms suffer from high propagation loop loss, LNOI platforms have high MZI propagation loss, and setups with off-chip loops experience high coupling loss for each loop. 
Further detailed comparisons on loss analysis can be found in the Supp. Mat. D.

\section{Evolution of Spin-Boson Model}
\begin{figure}[t]
\centering
\includegraphics[width=0.9\columnwidth]{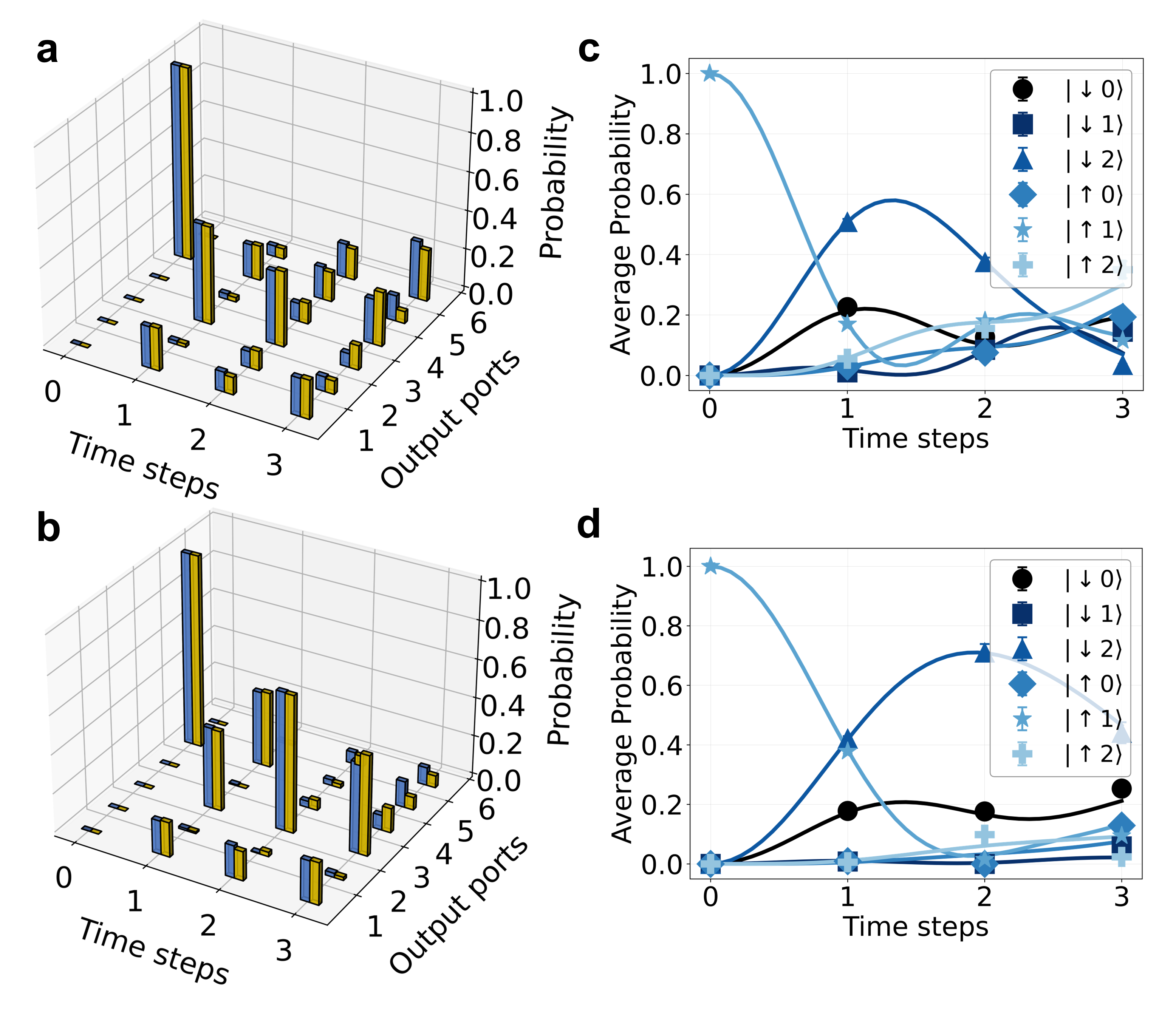}
\caption{(a,b) Probabilities for the six output channels at different time steps in experimental (yellow) and theoretical (blue) results. 
(c,d) Average probability, over twenty different realizations, of measuring a photon in each of the different channels at the different time steps: symbols for experimental data, and lines for theoretical predictions. The variance is generally smaller than the symbols. 
The Hamiltonian parameters used for (a,b) are $\epsilon=\hbar\omega=\lambda=h$ while $\epsilon=h/2$, $\hbar \omega=1.2h$, $\lambda=0.8h$ for (c,d).   
}
\label{fig:3}
\end{figure}

We now evaluate the performance of our Loop-QPC architecture in simulating the spin-boson model. As the initial condition, we choose a pure product state between the spin and the bosons, where the spin is in the excited state and the harmonic oscillator has zero excitations. We choose two scenarios with $\epsilon=\hbar\omega=\lambda=h$ and $\epsilon=h/2$, $\hbar \omega=1.2h$, $\lambda=0.8h$, for which the different terms of the Hamiltonian are similarly in value. This enables us to observe oscillatory and non-trivial dynamics without a large number of levels.  

The simulation of the system is corroborated by the probability of measuring a photon in an exit channel at the different time steps. Figure~\ref{fig:3} provides a comprehensive comparison between the experimental results and theoretical predictions for both scenarios. Specifically, Figs.~\ref{fig:3}(a,b) show histograms of the photon detection probabilities across output channels for several time steps, while Figs.~\ref{fig:3}(c,d) depict the temporal evolution of these probabilities. In Figs.~\ref{fig:3}(c,d), each data point represents the average of twenty experimental runs, and the error bars indicate the standard deviation.
Across both parameter regimes, we observe excellent agreement with theory, validating the accuracy and reliability of our Loop-QPC-based quantum simulation.

\begin{figure}[h]
\centering
\includegraphics[width=\columnwidth]{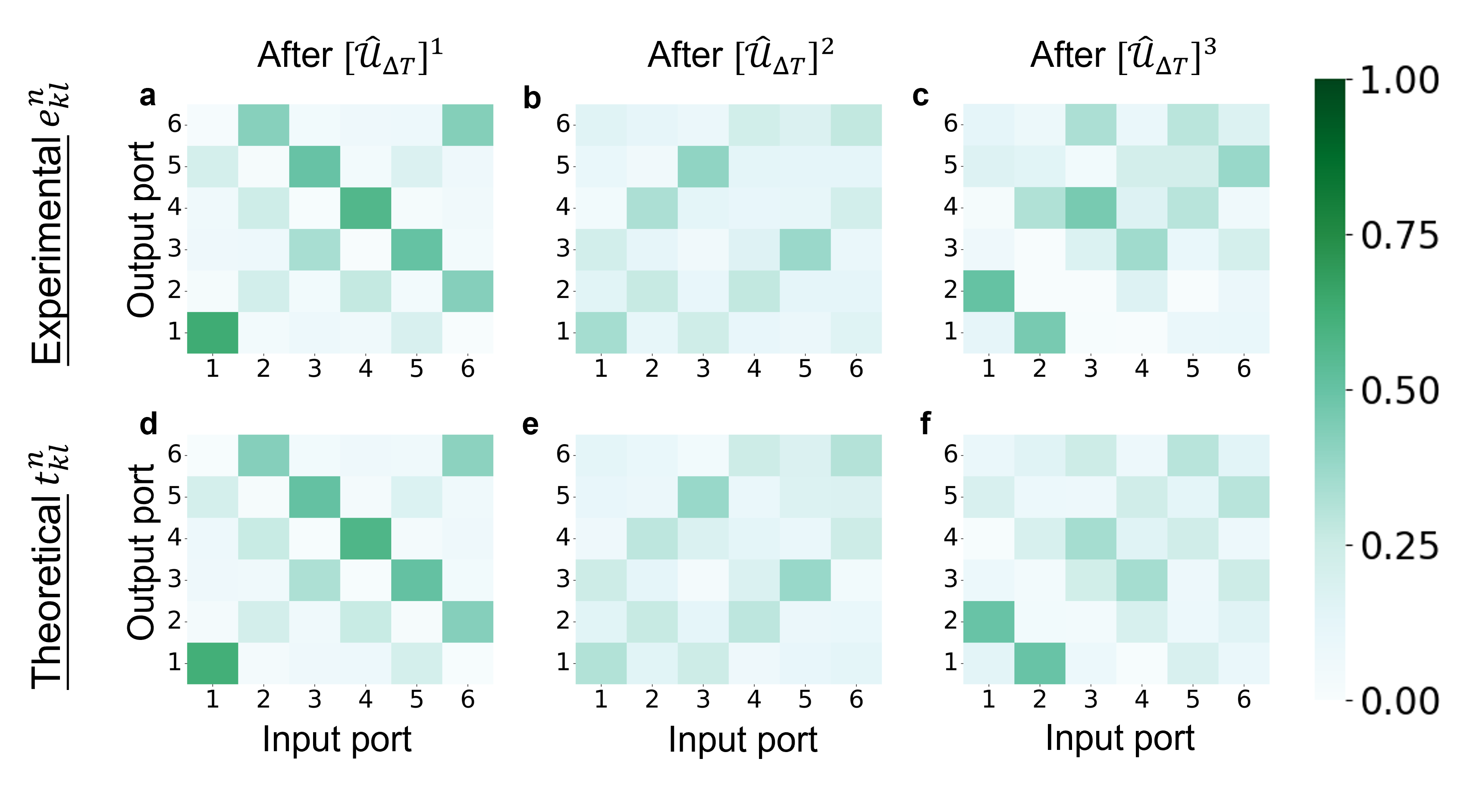}
\caption{Power matrices $e^n_{kl}$ and $t^n_{kl}$ for different input and output ports. (a)-(c) Experimental power matrices obtained from $\Ham_{SB}$ for $\epsilon=\lambda=\hbar\omega=h$, for different input and output ports. The matrix elements represent the transmittance at each of the output points, normalized along each input and output port, after zero (a), one (b), and two (c) loops. Darker shades of green indicate a higher transmittance for that input-output pair. (d)-(f) Theoretical results with the same parameter settings as in (a)-(c).} \label{fig:S1}
\end{figure}

To better quantify the accuracy of the method for generic inputs, we study the transmission across the chip for different loops. More precisely, we consider that a photon is sent through channel $k$ and we record when and in which channel it is measured. Then, for each of the different time steps $n$, we count the frequency of it being recorded in each output channel. We use two metrics: (i) the global fidelity \( F_n \), which yields \( F_1 = 0.978 \pm 0.005 \), \( F_2 = 0.943 \pm 0.016 \), and \( F_3 = 0.881 \pm 0.027 \) in our experiment; and (ii) the RMSE between experimental and theoretical power matrices \( e^n_{kl} \) and \( t^n_{kl} \) (where $l$ indicates the channel in which the photon was measured), respectively. Further details on fidelity and RMSE analysis can be found in the Supp. Mat. E and Fig. S1. The RMSE results are in Fig. \ref{fig:S1}, where panels (a)-(c) depict the experimental results and panels (d)-(f) the theoretical ones, showing a very good resemblance between theoretical predictions and experimental measurements.

\section{Scalability Analysis and Efficiency Comparison}
The number of modes of the chip determines the number of spins and bosonic levels that can be simulated from the spin-boson model. To simulate the model on a larger scale, we must consider the impact of increasing on-chip loss as the number of modes $N$ increases. Additionally, achieving higher efficiency in the Loop-QPC framework necessitates obtaining more evolution time steps $N_T$ within a single run. Thus, we discuss the potential to maximize $N$ and $N_T$ in integrated photonic circuit implementations.

\begin{figure}[t]
\centering
\includegraphics[width=0.9\columnwidth]{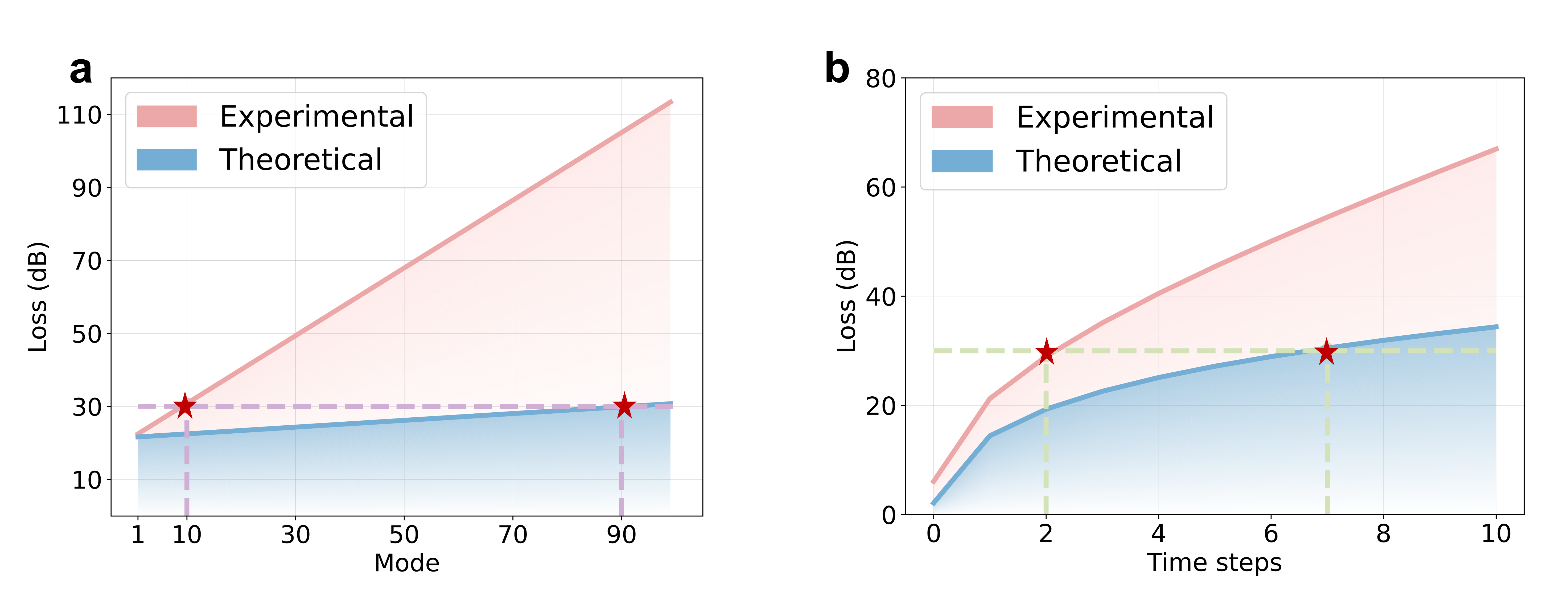}
\caption{Scalability analysis of Loop-QPC. (a) Estimated scalability of Loop-QPC in terms of system modes. Based on Eq.~\ref{eq:loss_theory} and near-term parameters (e.g., waveguide propagation loss $\alpha \sim 0.05$~dB/cm~\cite{bose2024anneal}), we calculate the maximum number of supported modes under a total loss of 30~dB. The star symbols indicate that current platforms can support at least 10 modes, while near-term low-loss implementations could scale up to 90 modes. (b) The potential of Loop-QPC in terms of the maximum achievable time steps. Based on parameters in Supp. Mat. D for experiment case, waveguide propagation loss ~\cite{bose2024anneal,tan2024ultra}, and other losses ($l_\text{others} \approx 3$) $d$B~\cite{zhang2024300}—we estimate the maximum number of time steps. The star symbols indicate that current implementations allow at least 2 time steps, while the near-term low-loss scenarios could support up to 7 time steps.}
\label{fig:scalability}
\end{figure}

The primary limitation in scaling $N$ arises from propagation loss $l_{chip}$ in Eq.\ref{eq:loss_theory}. Figure~\ref{fig:scalability}(a) (red line) shows our measured optical loss across varying mode numbers for a three-time-step identity matrix configuration, with $\alpha \approx 0.6$ dB/cm. The blue line shows a theoretical projection assuming an optimal propagation loss ($\alpha \approx 0.05$ dB/cm in ~\cite{bose2024anneal}). A total system loss of 30 dB is assumed, ensuring detectable photons with a minimum 10 dB signal-to-noise ratio at the SNSPD. These results indicate that our current platform supports QPC expansion up to $N = 10$, while under ideal low-loss conditions, scaling to $N = 90$ is feasible. This surpasses current state-of-the-art PICs~\cite{perez2025large,bao2023very,taballione202320}, confirming the scalability of our architecture.

Next, we evaluate the maximum achievable number of time steps. According to the loss calculations presented before, the total optical loss in Eq.\ref{eq:loss_theory} is governed by splitter loss, propagation loss, loop loss, coupling loss, and so on. Figure \ref{fig:scalability}(b) compares the total optical loss of our current implementation with near-term technology~\cite{bose2024anneal,tan2024ultra,zhang2024300}. Under optimal conditions, our model predicts achieving eight evolution step results (seven time steps), also assuming a 30 $d$B threshold. This result suggests that our method could provide around eight times higher efficiency compared to traditional approaches~\cite{sparrow2018simulating}. With further advancements in material science, fabrication techniques, and experimental equipment, our Loop-QPC design could enable even more time steps, highlighting its significant scalability.

In integrated quantum photonic platforms, most of the latency and energy consumption is due to the measurement system and the (classical) computation of the chip calibration parameters to implement the unitary evolution. For the Loop-QPC, measurement times only need to be performed once, while for the conventional photonic chips, one needs a time that increases linearly with the number of loops. This extra energetic cost becomes even more significant as the size of the photonic chips increases. In detail, the time required for computing unitary decompositions using Clements’ method naively goes as $T_{\text{com}}\propto N^4(N-1) t_c$ (detail in Supp. Mat. F), where $t_c$ denotes the average time per floating-point operation on a digital computer. The digital-to-analog operation time for MZI modulation is denoted by $T_{\text{da}}$, and the analog-to-digital conversion time for the SNSPDs is $T_{\text{ad}}$. The intrinsic optical computation time on-chip (from photon injection to detection) is negligible ($<4 \mathrm{~us}$). The time required for correlation measurement in SNSPD is $T_{\text{pd}}$. As the conventional methods do not have more losses at larger simulation times, their effective detection time is given by $\tilde{T}_{\text{pd}}=\sum_{n=0}^{N_T} T_{\text{pd}}/K^n $, where $K = 10^{L / 10,\mathrm{dB}}$ is the loss compensation factor and $L$ is the additional loss at each time step. In contrast, Loop-QPC executes all $N_T$ time steps sequentially on the same chip. Therefore, the total latency for Loop-QPC ($T_\text{lat}$) and for conventional approaches as in~\cite{sparrow2018simulating} ($\tilde{T}_\text{lat}$) for an $N_T$-step simulation are
\begin{align} T_\text{lat}=&T_{\text{com}}+T_{\text{da}}+T_{\text{ad}}+T_{\text{pd}}, \nonumber \\ 
\tilde{T}_\text{lat}=& N_T \cdot\left(T_{\text{com}}+T_{\text{da}}+T_{\text{ad}}\right)+\tilde{T}_{\text{pd}}. \end{align} 
$T_{\text{pd}}$ and $T_{\text{com}}$ are the dominant terms in experimental latency, and they increase with chip size $N$, thus allowing lower latency in the Loop-QPC approach.

Energetically, both Loop-QPC and conventional approaches involve on-chip power from MZIs and dominant off-chip power consumption from components such as lasers, DAC/ADC modules, and detectors. Similar to latency, as the total runtime of the system is reduced by approximately a factor of $N_T$, the overall energy consumption is also reduced by a similar factor. Also, Loop-QPC can achieve significant energy reduction, especially in regimes with large $N$ or small $K$.

\section{Conclusions and Outlook}
We have implemented a multi-step unitary evolution on a Loop-QPC platform, made possible by the inclusion of a programmable cycle-or-measure circuit that routes photons between processing and measurement. This mechanism allows us to simulate different time steps of the evolution in a single run, avoiding the need to recalibrate the chip to simulate different time steps or reconstruct intermediate states through quantum tomography, both of which reduce experimental overhead and increase efficiency. Our current demonstration focuses on unary encoding with single-photon states. The loop-based photonic architecture enables resource-efficient simulation of multi-step quantum dynamics~\cite{sparrow2018simulating}. This structure may also be extended to dual-rail and multi-photon regimes to access larger Hilbert spaces and explore computationally challenging problems~\cite{feng2023chip,spring2013boson}. Beyond exponential complexity regimes, some quantum applications~\cite{maraviglia2022photonic, smirne2010initial, somhorst2023quantum}  still benefit from efficient multi-time-step probing. These features suggest that our design could serve as a scalable tool for photonic simulation and offer a potential pathway toward studying many-body quantum systems.
Furthermore, our Loop-QPC architecture enables demonstrable savings in both time and energy. In conventional photonic approaches, the dominant resource costs stem from repeated unitary decompositions and measurements at each time step. The former scales with the number of steps $N_T$ and grows rapidly with circuit size $N$. In contrast, Loop-QPC performs all time steps within a single run, which eliminates redundant calibration and minimizes readout operations. It thus opens the way to an efficient and scalable solution for quantum simulations. 

Future developments would require a larger number of modes on the chips and of possible loops. Our calculation analysis shows that Loop-QPC remains within acceptable experimental limits even for state-of-the-art photonic integrated circuits.  Furthermore, comparing the total optical loss of our current implementation with the near-term technology, an increase from 3 to 8 loops is within reach. With the reduction of losses (e.g., through ultra-low propagation loss waveguides~\cite{bose2024anneal}, the use of high-speed photonic switching technologies~\cite{wang2018integrated}), and improved tuning of the chip parameters,  
Loop-QPC could thus open the way to simulate complex dynamics, including open dynamics, either via collisional models \cite{erbanni2023simulating, mi2024stable} or finite-size baths \cite{xu2022typicality}.

\subsection*{Funding}
D.P. and K.L.C. acknowledge support from the National Research Foundation, Singapore, under its QEP2.0 program (NRF2021-QEP2-02-P03 and NRF2022-QEP2-02-P16). K.L.C. acknowledges joint support from the National Research Foundation, Singapore, and the Ministry of Education under EEE and CQT. Z.H. acknowledges support by National Natural Science Foundation of China (62505228), Chenguang Program of Shanghai Education Development Foundation and Shanghai Municipal Education Commission (24CGA19), Shanghai Science and Technology Innovation Action Plan Fundamental Research Program in Integrated Circuit (25JD1406000), and Fundamental Research Funds for the Central Universities.

\subsection*{Acknowledgment}
Y.C.Z., H.Z., and A.B. jointly conceived the idea. Y.C.Z., H.Z., and L.X.W. designed the chip and built the experimental setup. Y.C.Z. performed the experiments. H.Z. assisted with the setup and experiment. D.P., K.L.C., and X.D.J. assisted with the theory. All authors contributed to the discussion of experimental results. K.L.C., D.P., and A.Q.L. supervised and coordinated all the work. Y.C.Z., D.P., and R.E. wrote the manuscript with contributions from all co-authors.

\subsection*{Disclosures}
The authors declare no conflicts of interest.

\subsection*{Data availability}
Data underlying the results presented in this paper are available from the corresponding authors upon reasonable request.

\subsection*{Supplemental document}
See Supplement Materials for supporting information.

\nolinenumbers

\setcounter{figure}{0}     
\renewcommand{\thefigure}{A\arabic{figure}}

\bibliography{mybib}

\end{document}